\documentclass{osa-article}
\journal{osajournal}
\articletype{Research Article}

\usepackage[utf8]{inputenc} % set input encoding (not needed with XeLaTeX)
\usepackage{graphicx}
\usepackage{caption}
\usepackage{float}
\usepackage{wrapfig}

\usepackage{graphicx} % support the \includegraphics command and options
\usepackage{amsmath}
\usepackage{bm}
\usepackage{mathtools}

\usepackage{booktabs} % for much better looking tables
\usepackage{array} % for better arrays (eg matrices) in maths
\usepackage{paralist} % very flexible & customisable lists (eg. enumerate/itemize, etc.)
\usepackage{verbatim} % adds environment for commenting out blocks of text & for better verbatim
\usepackage{subfig} % make it possible to include more than one captioned figure/table in a single float
\newcommand{\prob}{\mathrm{pr}}
\newcommand{\Prob}{\mathrm{Pr}}

\newcommand{\ent}{\mathrm{H}}
\newcommand{\smi}{\mathrm{I_S}}

\newcommand{\pe}{\mathrm{P_e}}
\DeclarePairedDelimiterX{\infdivx}[2]{(}{)}{%
  #1,#2%
}
\newcommand{\kld}{\ensuremath{\mathrm{KL}\infdivx}}
\newcommand{\bd}{\ensuremath{\mathrm{BD}\infdivx}}

\begin{document}

\title{X-ray measurement model incorporating energy-correlated material variability and its application in information-theoretic system analysis}

\author{Yijun Ding,\authormark{1,*} and Amit Ashok,\authormark{1,2}}

\address{\authormark{1}Wyant College of Optical Sciences, University of Arizona, Tucson AZ 85721\\
\authormark{2}Department of Electrical and Computer Engineering, University of Arizona, Tucson AZ 85721
}

\email{\authormark{*}dingy@email.arizona.edu} %% email address is required

%%%%%%%%%%%%%%%%%% abstract %%%%%%%%%%%%%%%%
% [use \begin{abstract*}...\end{abstract*} if exempt from copyright]

\begin{abstract}
Extending our prior work, we propose a multi-energy X-ray measurement model incorporating material variability with energy correlations to enable the analysis and exploration of the performance of X-ray  imaging and sensing systems. Based on this measurement model, we provide analytical expressions for bounds on the probability of error, $\pe$, to quantify the performance limits of an X-ray measurement system for binary classification task. We analyze the performance of a prototypical X-ray measurement system to demonstrate the utility of our proposed  material variability measurement model. 
\end{abstract}

%%%%%%%%%%%%%%%%%%%%  Introduction  %%%%%%%%%%%%%%%%%%%%%%%
\section{Introduction}
\indent  Imaging and sensing based on X-ray attenuation is commonly used to non-destructively discriminate materials in security screening, medical imaging and industrial inspection \cite{wells2012review, hsieh2009computed, hanke2008x}. An X-ray measurement model with accurate statistics of the measurement data is desirable for many purposes such as, evaluating system performance, developing detection classification algorithms, and optimization of object-reconstruction algorithms.

\indent The performance of attenuation-based X-ray imaging and sensing systems is limited by at least two fundamental factors in the measurement data: shot noise and inherent material variability \cite{barrett1990objective}. The shot noise stems from the randomness in the generation, attenuation and detection of X-ray photons. The material variability, arising from the inherent fluctuations in material composition and density, limits the discrimination of materials. Therefore, a rigorous evaluation of an X-ray imaging or sensing system must take into account both shot noise and material variability. 

\indent An X-ray measurement model that considers only the shot noise has been analyzed in the context of material-discrimination applications  by Huang et.\ at\ and Lin et.\ al.\ \cite{huang2015information, lin2016information}. Recently, Masoudi et.\ at.\ \cite{masoudi2018x} improved the model by incorporating material variability under the assumption of energy (statistical) independence. However, the energy-correlations are intrinsic in X-ray attenuation and cannot be ultimately ignored. In this paper, we propose an X-ray measurement model that considers energy-correlated material variability as well as shot noise. 

\indent  Many imaging systems are used to perform binary-classification tasks. For example, a luggage scanner at an airport checkpoint is used to determine whether a luggage bag contains threat material or not; mammography is often used to classify tumor-present or tumor-absent; in industrial non-destructive evaluation, radiographs of parts are used to examine the existence of defects. An objective assessment of an system must take into account the task of the system \cite{neifeld2007task, barrett1990objective}. 

\indent Task-specific information (TSI) \cite{neifeld2007task}, which is the information content relevant to the task, is commonly used as an objective assessment metric. For classification tasks, Shannon mutual information ($\smi$) \cite{cover2012elements} is a natural choice as an information-theoretic metric, because it can be used to bound probability of classification error ($\pe$) \cite{fano1961transmission, cover2012elements,kovalevskij1967problem, tebbe1968uncertainty, hu2016optimization}. Although $\smi$ is expensive to compute for non-trivial distributions, we are able to derive closed-form expressions for bounds on $\smi$ and bounds on $\pe$ for mixture distributions with the help of a recent work \cite{kolchinsky2017estimating}. 

\indent This paper is organized as follows. We derive the measurement model in Sec~\ref{sec:meas_model}. Sec~\ref{sec:metric} reviews $\smi$, $\pe$ for binary-classification tasks and derives closed-form expressions for bounds on $\smi$ and bounds on $\pe$. In Sec~\ref{sec:result}, as an example application, we apply the model to a simplified luggage-scanner and present the simulation results. Sec~\ref{sec:discussion} discusses the advantages and drawbacks of the model. Sec~\ref{sec:conclusion} provides a succinct conclusion.

%%%%%%%%%%%%%%%%%%%%%%%% model %%%%%%%%%%%%%%%%%%%%%%
\section{Measurement model} \label{sec:meas_model}
 
The measurement model relates the data to the object and describes the statistical properties of the data. A general form of the measurement model can be written as
\begin{equation}
	{\bm g} = \mathcal{H}{\bm f} + {\bm n} = \mathcal{H}(\bar{\bm f}+\Delta{\bm f}) + {\bm n},
\end{equation}
where ${\bm g}$ is the data, $\mathcal{H}$ describes the system, $\bm f$ is the object, and $\bm n$ is the system noise, $\bar{\bm f}$ is the ensemble mean of the object, and $\Delta{\bm f}$ describes the material variability.

\subsection{X-ray attenuation coefficient $\mu$}
\indent In the energy range commonly used for X-ray transmission imaging, the interaction between X-ray photons and the medium can be categorized into the following three processes: photoelectric absorption, Compton scattering and coherent (Rayleigh) scattering. In photoelectric absorption, a photon disappears and the energy of the photon transfers to an electron in the material; in Compton scattering, an X-ray photon is deflected and transfers a portion of its energy to an electron; and in coherent scattering, an X-ray photon is deflected, but retains its energy. The strength of each interaction process can be characterized by the energy of the photon $E$, the atomic number of the medium $Z$ and the density of the medium $\rho$. For a material with fixed $Z$ and $\rho$, the X-ray attenuation coefficient $\mu(E)$ is a function of the X-ray energy $E$. Variations in $\mu(E)$ due to variability in $Z$ and $\rho$ demonstrate intrinsic energy correlation.

\indent For multi-element compounds and mixtures, the attenuation coefficient is
\begin{equation}
	\mu(E) = \rho\sum_{c}\frac{w_c}{\rho_c}\mu_c(E) 
\end{equation}
where $\rho$ represents the density of the medium, $w_c$ is the weight fraction of the $c^{th}$ element in the compound or mixture, $\mu_c(E)$ and $\rho_c$ are the attenuation and the density of the $c^{th}$ element. When there is variability inherent in the description of a material, which may stem from density fluctuations, composition variations and packaging differences, the attenuation coefficient $\mu(E)$ becomes a random process.

\indent To consider material variation, we assume $\mu(E)$ is a Gaussian random process with covariance function $\Sigma_{\mu}(E,E')$. For any set of energy $E_1$, $E_2$, ..,$E_R$, the random variable $\mu(E_1)$, $\mu(E_2)$, ..., $\mu(E_R)$ are joint Gaussian random variables. If we denote the set of random variables $\mu(E_i)$ by the vector $\bm\mu$, whose mean is $\bm\mu_{0}$ and the covariance matrix is $\Sigma_{\mu}$,
then the joint probability density is

\begin{equation}
	\prob(\bm\mu) = \mathcal N(\bm\mu|\bm\mu_{0}, \Sigma_{\mu}) = \frac{\exp[-\frac{1}{2}(\bm\mu-\bm\mu_{0})^T\Sigma_\mu^{-1}(\bm\mu-\bm\mu_{0})]}{\sqrt{2\pi^R |\Sigma_{\mu}|}},
\end{equation}
where $|\Sigma_{\mu}|$ is the determinant of $\Sigma_{\mu}$, $(\cdot)^T$ is the transpose of matrix $(\cdot)$, $R$ is the length of the vector $\bm\mu$. When $\Sigma_{\mu}$ is singular, the Gaussian distribution turns into a Dirac delta function in the corresponding dimension.  

\subsection{X-ray attenuation $\tau$}

\begin{figure} [H]
\centering
\includegraphics[width=0.4\linewidth]{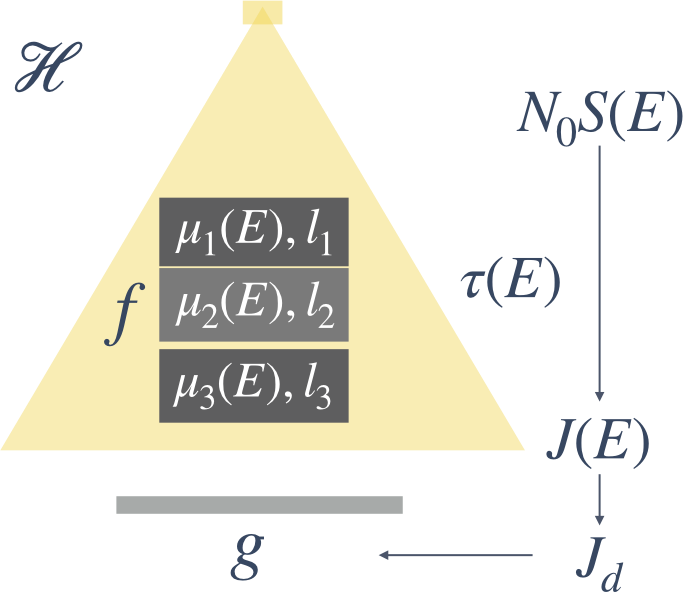}
\caption{Illustration of X-ray penetrating multiple objects.}
\label{fig:illustration_system}
\end{figure}

\indent Now consider a beam of X-ray penetrating multiple items, as illustrated in Figure~\ref{fig:illustration_system}, where each item contains one material. If there are a total of $N_{it}$ items between the X-ray source and the detector, the total attenuation along the path is
\begin{equation}
	\tau(E) = \int \mu(E,l)~\mathrm{dl} = \sum_{t=1}^{N_{it}}\mu_{t}(E)\,l_{t},
\end{equation}
where $t$ is the index for items along the X-ray path, $l_{t}$ is the length of the material along  the path, and $\mu_{t}(E)$ is the attenuation profile of the material in the ${t}^{th}$ item. In the following discussion, we refer to $\tau(E)$ as the total attenuation. 

\indent Under the assumption that the fluctuations of $\mu(E)$ in two different items are independent, $\tau(E)$ is a Gaussian random processes, since it is the sum of independent Gaussian random processes, $\mu_t(E)$. For a set of energies $E_1$, $E_2$, ..,$E_K$, the set of total attenuation $\bm\tau$ follows a multivariate normal distribution
\begin{equation}
	\prob(\bm\tau) = \mathcal N(\bm\tau|\bm\tau_{0}, \Sigma_{\tau}) = \frac{\exp[-\frac{1}{2}(\bm\tau-\bm\tau_{0})^T\Sigma_\tau^{-1}(\bm\tau-\bm\tau_{0})]}{\sqrt{2\pi^R |\Sigma_\tau|}},
\end{equation} 
where $R$ is the length of the vector $\bm\tau$ and
\begin{equation}
	\bm\tau_{0}=  \sum_{t=1}^{N_{it}}\bm\mu_{0,t}\, l_{t},
\end{equation}

\begin{equation}
	\Sigma_{\tau} =  \sum_{t=1}^{N_{it}} \Sigma_{\mu,t}\, l_{t}^2.
\end{equation}

\subsection{Beer's law}
\indent An illustration of the propagation of X-ray photons starting from the X-ray source, through the object, and ending with photon detection is shown in Figure~\ref{fig:illustration_system}. The X-ray attenuation follows beer's law,
\begin{equation}
	J(E) = \frac{N_0S(E)}{t} e^{-\tau(E)},
	\label{eq:beers_law}
\end{equation}
where $J(E)$ is the mean spectral flux incident on a detector element at energy $E$, $N_0S(E)/t$ is the source spectral flux, $\tau(E)$ is the total attenuation as a function of $E$, and $t$ is the exposure time. More specifically, $N_0$ is the number of photons emitted from the X-ray tube in the solid angle extended by a detector element over the exposure time $t$, and $S(E)$ is the normalized x-ray source spectrum. The units of $J(E)$ and $S(E)$ are (s$\cdot$keV)$^{-1}$ and keV$^{-1}$, respectively. We denote $\tau(E) = \tau_0(E) + \Delta\tau(E)$, where $\Delta\tau(E)$ is fluctuation or perturbation in attenuation around $\tau_0(E)$, and $J_0(E) = N_0 S(E) e^{-\tau_0(E)}/t$. When the material variance (perturbation) is small, or more specifically, when $\Delta\tau(E)\ll 1$, we can approximate $e^{-{\Delta\tau(E)}}$ with the first two terms in the Taylor expansion, resulting in:
\begin{equation}
	J(E) \approx J_0(E) - J_0(E){\Delta\tau(E)}.
	\label{eq:beers_law_linear}
\end{equation}
The remainder of the approximation can be bounded by 
\begin{equation}
	|R_2(E)| < \frac{J_0(E)|\Delta\tau(E)|^3}{6}.
\end{equation}
If a requirement on the remainder is to be less than $1\%$ of $J_0(E)$, $|\Delta\tau(E)|$ should be less than 0.39. One can always guarantee such a requirement by splitting a material with large variations into multiple materials with smaller variations. For example, 40$\%$ sugar water with sugar content varying from 20$\%$ to 60$\%$ can be split into two materials with sugar content varying from 20$\%$ to 40$\%$ and 40$\%$ to 60$\%$, respectively. 

\indent In the limit of small variability, the two-term approximation of $J(E)$, as defined by Equation~(\ref{eq:beers_law_linear}), follows a normal distribution when $\tau(E)$ follows a normal distribution. Therefore, $\bm J$, which is $J(E)$ at a set of energies, approximately follows a normal distribution:
\begin{equation}
	\prob(\bm J) = \mathcal N(\bm J | {\bm J_0},\Sigma_J).
\end{equation}
The mean and covariance matrix are
\begin{equation}
 {\bm J}_0 = \frac{N_0\bm S}{t} \odot e^{-\bm\tau_0},
\end{equation}
and
\begin{equation}
	\Sigma_J =  (\bm J_0\bm J_0^T) \odot \Sigma_\tau,
	\label{eq:cov_IE}
\end{equation}
where $\odot$ is element-wise multiplication. 

\subsection{Detector response and energy binning}\label{ssec:detector_binning}
\indent If the energy response of the detector is linear, one can express the mean photon count collected in the $m^{th}$ energy bin as
\begin{equation}
	J_m = t\int_0^\infty J(E) D_m(E) \mathrm{dE},
\end{equation}
where $D_m(E)$ is the detector response of the $m^{th}$ energy bin to a photon with energy $E$ and $t$ is the exposure time. 

\indent With a total of $M$ energy bins, the mean photon count $\{J_1, J_2, ..., J_M\}$ after energy binning can be represented by a vector $\mathbf J_d$, and
\begin{equation}
\mathbf J_d = t \mathcal D\,J(E),
\end{equation}
where $\mathcal D$ is the detector response operator and the subscript $d$ denotes detector.

\indent When $\mathcal D$ is a linear operator and $J(E)$ is a Gaussian random process, $\bm J_d$ follows a normal distribution. The mean and variance are
\begin{equation}
	\mathbf J_{d0} = t D\mathbf J_0
	\label{eq:mean_photon_count}
\end{equation}
and
\begin{equation}
	\Sigma_{J_d} = t^2 D\Sigma_J D^T,
	\label{eq:cov_photon_count_matvar}
\end{equation}
where $D$ is the matrix form of the operator $\mathcal D$ for a set of $R$ energies. 

\subsection{Shot noise}

\indent In attenuation-based X-ray imaging, the data collected is often the number of detected X-ray photons. Photon counting intrinsically introduces shot noise, hence the number of X-ray photons detected in one energy bin, $g$, follows a Poisson distribution:
\begin{equation}
\Prob(g | J_{d}) = {\mathcal Poiss} (g | J_{d}) =\frac{(J_{d})^{g}e^{-J_{d}}}{g!},
\end{equation}
where $\mathcal Poiss$ indicates a Poisson distribution and $J_d$ is the mean photon count in the energy bin. 

\indent When the mean photon count is relatively large (i.\ e.\ more than say, 10 photons), the Poisson distribution can be approximated by a Gaussian/normal distribution \cite{barrett1996radiological}. Denoting the continuous variable $x = g$,
\begin{equation}
\prob(x | J_{d}) \approx \mathcal N(x; J_{d}, J_{d}) = \frac{\exp[-(x - J_{d})^2/(2J_{d})]}{\sqrt{2\pi J_{d}}}.
\end{equation}
When the mean photon count $J_{d}\geq 100$, the error introduced by the Gaussian approximation is less than $4\%$ for $x=\bar{x}\pm\sigma(x)$, where $\bar{x}$ and $\sigma(x)$ are the mean and standard deviation of $x$, respectively. A derivation of the percentage error is given in Appendix~B. 

\subsection{Combined model}

\indent In our measurement model, we combine statistics of the shot noise and the statistics of the energy-correlated material variability. Thus, the covariance of the measurement data is a summation of the covariance matrices corresponding to the shot noise and the material variability. More specifically, if we denote the continuous data vector as $\bm x$, the probability density function (PDF) of $\bm x$ with mean photon counts, $\bm J_{d0}$, and a covariance matrix induced by material-variation, $\Sigma_{J_d}$, is
\begin{equation}
		\prob(\bm x| \bm J_{d0}, \Sigma_{J_d}) \approx \mathcal N(\bm x;\, \bm J_{d0}, \,\Sigma_{J_d} + \text{diag}(\bm J_{d0})).
		\label{eq:one_bag_data}
\end{equation}
A detailed derivation is provided in Appendix~C. 

\indent We define the combined data over all detector pixels as $\bm g$. The probability distribution function of $\bm g$ is the joint distribution of all $\bm x_n$, 
\begin{equation}
	\prob(\bm g) = \prob_{\bm x_1, \bm x_2, ... \bm x_N}(\bm x_1, \bm x_2, ... \bm x_N),
\end{equation}
where $n$ is  the detector-pixel index and $N$ is the number of detector pixels. The joint distribution is determined by the geometry of the object and setup of the system. 

\indent Now, consider an ensemble of K objects that are examined by an X-ray system. Then the data representing the ensemble of objects can be described by the following mixture distribution:
\begin{equation}
	\prob(\bm g|\bm a) = \sum_{i=1}^{K} a_i \prob_i(\bm X),
\end{equation}
where $a_i$ is the probability of the $i_{th}$ object occurring in the ensemble, $K$ is the number of objects in the ensemble, $\sum a_i = 1$, and $ \prob_i(\bm X)$ is the probability distribution function of measurement data $X$ when the $i^{th}$ object is imaged. Note that we have used the discretized data $\bm g$ and the continuous variables $\{\bm x_1, \bm x_2, ... \bm x_N\}$ interchangeably.

%%%%%%%%%%%%%%%%%%%  metric %%%%%%%%%%%%%%%%
\section{TSI for binary classification tasks} \label{sec:metric}

\indent Shannon mutual information, $\smi$, has long been used as a metric to quantify the task-specific fidelity of a measurement with respect to classification tasks \cite{neifeld2007task}. This is because $\smi$ is related to the error probability, $\pe$ through Fano's inequality \cite{fano1961transmission} and Kovalevskij's inequality \cite{kovalevskij1967problem, tebbe1968uncertainty, hu2016optimization}. In this section, we provides a brief summary of the relation between $\smi$ and $\pe$ and provides closed-form expressions for bounds on $\smi$ and $\pe$ defined on our X-ray measurement model.

\indent The system performance is object dependent. Properties of the object, such as the size, the material and the geometry, affect the distribution of the data and hence the difficulty of the classification task. To reduce the dependence on test objects, a general assessment of a system should consider a large ensemble of objects. In the following discussion, we consider an ensemble of $K$ objects that consists of $K_1$ objects in the first class and $K_2$ objects in the second class. The probability of the $i^{th}$ object in the ensemble is $a_i$. The probabilities of the two class labels are $P_1$ and $P_2$, where $P_1 + P_2 = 1$ for binary classification.

\subsection{Bounds on $\smi$}
The $\smi$ is defined as
\begin{equation}
	\smi(\mathbf g; C) = \ent(\bm g) - \sum_{c=1}^2 P_c\ent(\bm g|C=c),
\end{equation}
where $\ent(\bm g)$ is the Shannon entropy of the distribution of the measured data, and $\ent(\bm g|C=c)$ is the entropy of the conditional distribution of the measured data given the class $C$.

\indent When the data is a mixture distribution, the $\smi$ between data and class label has no closed-form expressions. However, $\smi$ can be bounded by bounds of the entropy  \cite{kolchinsky2017estimating}, 
\begin{equation}
\begin{split}
	\smi(\mathbf g;C) &\geq \hat \ent_{BD}({\bf g}) - \sum_{c=1}^2 P_c \hat \ent_{KL}({\bf g} |C=c), \\
	\smi(\mathbf g;C) &\leq \hat \ent_{KL}({\bf g}) - \sum_{c=1}^2 P_c\hat \ent_{BD}({\bf g} |C=c),
\end{split}
\label{eq:bounds_Is}
\end{equation}
where $\hat\ent_{BD}({\bf g})$ and $\hat\ent_{KL}({\bf g})$ are lower and upper bounds on entropy based on pair-wise Bhattacharyaa distance (BD) and pair-wise Kullback-Leibler (KL) divergence, respectively. More specifically, the bound on entropy based on either divergence is given by
\begin{equation}
\hat \ent_{D}(\mathbf g) = \sum_{i=1}^{K} a_i \ent({\prob}_i)  - \sum_{i=1}^{K} a_i \ln\sum_{j=1}^{K} a_j\exp(-\mathrm{D}({{\prob}_i, {\prob}_j})),
\end{equation}
where ${\prob}_i$ is the PDF of the data $\bm g$ if the $i^{th}$ bag is measured and $\mathrm{D}(p,q)$ can be either Bhattacharyaa distance or KL divergence, which are defined by
\begin{equation}
\begin{split}
	\bd{p}{q} & = -\ln\int\mathrm{dx}~\sqrt{p(x)\,q(x)}, \\
	{\text and} \quad \kld{p}{q} &= \int\mathrm{dx}~p(x)\ln\frac{p(x)}{q(x)},
	\end{split}
\end{equation}
respectively. 

\indent An upper bound on $\smi$ based on pair-wise KL divergence and a lower bound based on pair-wise BD are provided in Appendix~A. The minimum of the two upper bounds and the maximum of the two lower bounds can be used as the tighter version of the bounds on $\smi$.

\subsection{Bounds on $\pe$}
\indent Starting from $\smi$, the Fano's inequality \cite{fano1961transmission} provides a lower bound on $\pe$ for binary classification, as following
\begin{equation}
	P_e \geq h_b^{-1}[\ent(C) - \smi(\mathbf g;C)],
\end{equation}
where $h_b(x) = -x\log_2(x) - (1-x)\log_2(1-x)$ is the binary entropy function, $h_b^{-1}(\cdot)$ is the inverse function of $h_b(\cdot)$, and $\ent(C)=h_b(P_1)$ is the Shannon entropy of the class label $C$. More specifically, one can calculate $P_e$ by placing the value $\ent(C) - \smi(\mathbf g;C)$ on the left side of the binary entropy function and solving for $x$. When $P_e\ll1$, $\ent(C) - \smi(\mathbf g;C)\approx-\pe\log{\pe}$ and hence is on the same order of magnitude as $P_e$.

\indent An upper bound on binary classification errors $\pe$, which is tighter than Kovalevskij's inequality, has been reported recently \cite{hu2016optimization},
\begin{equation}
	P_e \leq {\text{min}} \left\{P_{min},\, f_{ub}^{-1}[\ent(C) - \smi(\mathbf g; C)]\right\},
	\label{eq:bounds_pe2}
\end{equation}
where $P_{min}$ is min$\{P_1, P_2\}$, and $f_{ub}(x)$ is an upper bound function defined by
\begin{equation}
	f_{ub}(x) = -P_{min}\log_2\frac{P_{min}}{x+P_{min}} - x\log_2\frac{x}{x+P_{min}},
\end{equation}
and $f_{ub}^{-1}(\cdot)$ is the inverse function of $f_{ub}(\cdot)$.

\subsection{Closed-from expressions for pair-wise BD and KL divergence}

\indent If we assume that the measurement data at different pixels are statistically independent with each other, the PDF of the measurement data becomes a product of the PDFs of the data measured at all pixels:
\begin{equation}
	\prob(\bm g) = \prod_{n=1}^{N}\prob(\bm x_{n}\,|\,\bm J_{d0,n}, \Sigma_{J_d,n}) \approx  \prod_{n=1}^{N} \mathcal N(\bm x_{n}\,; \bm J_{d0,n}, \Sigma_{J_d,n}),
\end{equation}
where $n$ is the detector-pixel index, and $N$ is the total number of detector pixels.

\indent Calculation of bounds on $\smi$ and $\pe$ require pair-wise Bhattacharyaa distance and pair-wise KL divergence. To simplify notation, we define
\begin{equation}
	\Delta \mathbf J_{n} = \mathbf J_{d0,n,i} - \mathbf J_{d0,n,j}, 
	\label{eq:delta_I}
\end{equation}
\begin{equation}
	\Sigma_{n,i} = \Sigma_{J_d,n,i}+ \text{diag}(\bm J_{d0,n,i}),
	\label{eq:sigma_i}
\end{equation}
and
\begin{equation}
	\Sigma_{n} = \Sigma_{n,i}+ \Sigma_{n,j}.
	\label{eq:sum_sigma}
\end{equation}

The analytical form of Bhattacharyya distance can be expressed as 
\begin{equation}
	\bd{\mathrm{\prob}_i}{\mathrm{\prob}_j} = \sum_{n=1}^N\left[ \frac{\Delta \mathbf J^T_{n}\Sigma_{n}^{-1}\Delta \mathbf J_{n}}{4} -\frac{\ln{|\Sigma_{n,i}\Sigma_{n,j}|}}{4}+\frac{\ln{|\Sigma_{n}|}}{2} -\frac{M\ln{2}}{2}\right];
\end{equation} 
and the analytical form of KL divergence can be expressed as
\begin{equation}
	\kld{\mathrm{\prob}_i}{\mathrm{\prob}_j} = \frac{1}{2}\sum_{n=1}^N \Delta \mathbf J^T_{n}(\Sigma_{n,j})^{-1}\Delta \mathbf J_{n} -\ln{|\Sigma_{n,i}|}+\ln{|\Sigma_{n,j}|} + \mathrm{tr}(\Sigma_{n,j}^{-1}\Sigma_{n,i}) - M,
\end{equation}
where $\mathrm{tr}(\cdot)$ is the trace of the matrix. 

%%%%%%%%%%%%%%%%%%%% Results %%%%%%%%%%%%%%%%%%%%%%%%%
\section{Illustrative System Study and Results} \label{sec:result}

\begin{figure}[H]
\centering
	\includegraphics[width=0.95\linewidth]{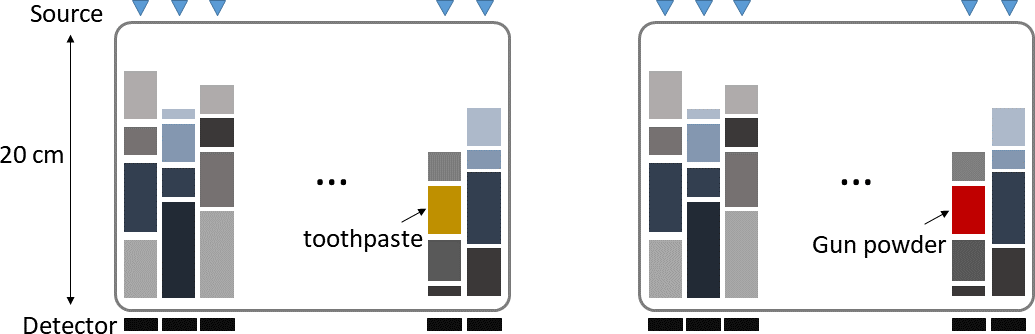}
	\caption{Illustration of the simulated X-ray luggage scanner.}
\label{fig:illustration_simu}
\end{figure}

\indent In this section, we apply our measurement model to study a simple X-ray measurement system for the task of material-based threat detection (i.\ e.\ a binary classification problem). The X-ray system, as illustrated in  Figure~\ref{fig:illustration_simu}, has 10 X-ray sources that produce parallel pencil-beams and 10 corresponding photon-counting energy-sensitive detector elements. A source with a tungsten target operating at 160 kVp is assumed, and the corresponding source spectrum was generated with SpekCalc \cite{poludniowski2009spekcalc}. The energy-sensitive detector can have 1, 2 and 3 energy bins. The bin edges are determined by balancing the photon count after attenuation. More specifically, the bin edges are [30, 160]~keV for one bin, [30, 70, 160]~keV for two bins, and [30, 60, 85, 160]~keV for three bins. The objects under inspection contain 10 vials of materials, and the location of each vial is along one parallel-beam X-ray path. Each vial contains 4 materials, which is randomly sampled from a library of materials. The lengths of the vials are randomly chosen between 0.5~cm to 20~cm.

\indent A material library and a variability model of the material composition has been previously developed \cite{masoudi2018x}. The material library contains 25 threat materials and 33 non-threat materials. Examples of threat materials are ammonium nitrite, hydrogen peroxide and gun powder; examples of non-threat materials are milk, toothpaste and polyethylene. The composition and variance of weight fractions of each material were determined based on industrial standards. The density variation was folded in either by varying the density $\rho$ or by adding air as a component. For each material, 1000 composition realizations were randomly generated; and for each material realization, an X-ray attenuation profile was computed based on the NIST XCOM database \cite{berger1998photon}. For each material, the mean and covariance of the 1000 attenuation profiles were used as the mean and the covariance of $\mu(E)$ in our model. The number of energy samples in $\bm \mu$ and the source spectrum is $R=180$.

\indent We simulate objects, where each object has 10 vials, in pairs and each pair consists of one object containing one threat material and one object containing no threat material. A threat object and a non-threat object in a pair share the same geometry (aka.\ vial lengths and materials) and are thus different by only one material. An illustration of an object pair is shown in Figure~\ref{fig:illustration_simu}. We simulate objects in pairs because of the following two reasons: (1) the system performance is dominated by objects that are located close to the class boundary in the data space; and (2) in general, the distance in the data space between two objects in a pair is closer than that between two random objects. 

\indent An ensemble of 160 bag-pairs was simulated. Equal prevalence of each class and equal probability of occurrence of each bag-pair were assumed for the calculation of the TSI measure. 

\indent There is no spatial correlation between data measured from different detector elements in this prototypical example. The closed-form expressions derived under the assumption of spatial independence were used to calculate the bounds on $\smi$ and $\pe$. First, we study the behavior of bounds on $\smi$ and bounds on $\pe$ for one, two and three energy bins. Second, we compare the following three measurement models in terms of bounds on $\pe$: (1) shot noise only, (2) material variation only, and (3) combined model. Lastly, we compare the energy-correlated measurement model and material-variation model with the energy-uncorrelated models in terms of lower bound on $\pe$. 

\subsection{Behavior of TSI metrics}
In this section, we present results calculated with the energy-correlated measurement model that incorporates both shot noise and material variation and illustrate the behavior of the TSI metrics.
 
\begin{figure}[H]
\centering
	\includegraphics[width=\linewidth]{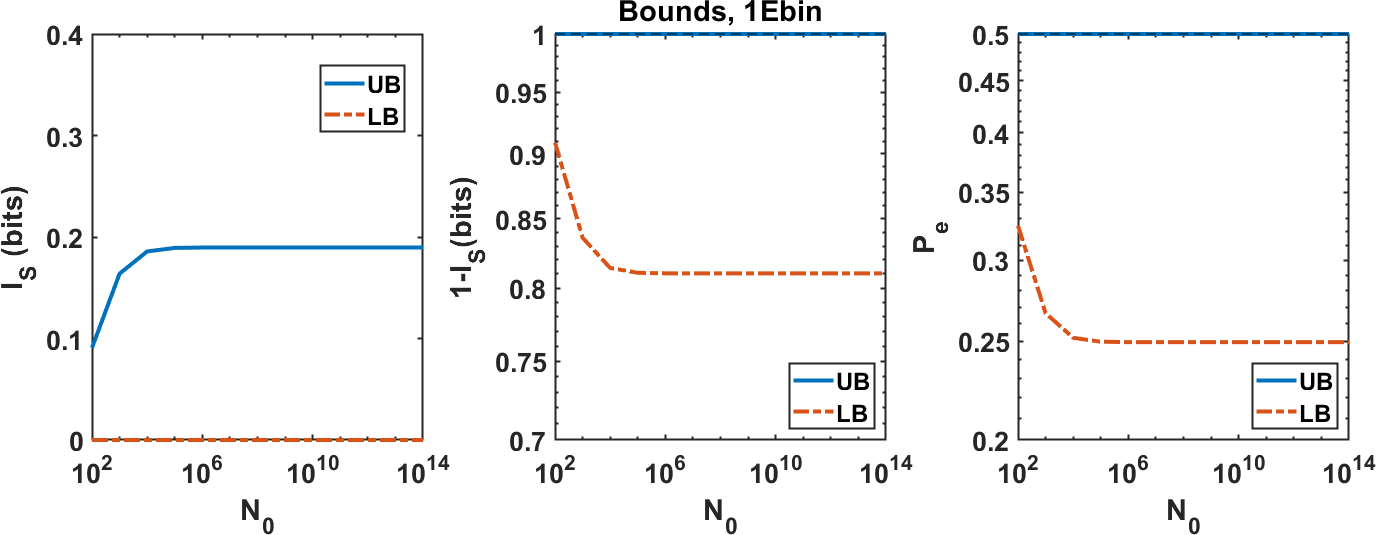}
	\includegraphics[width=\linewidth]{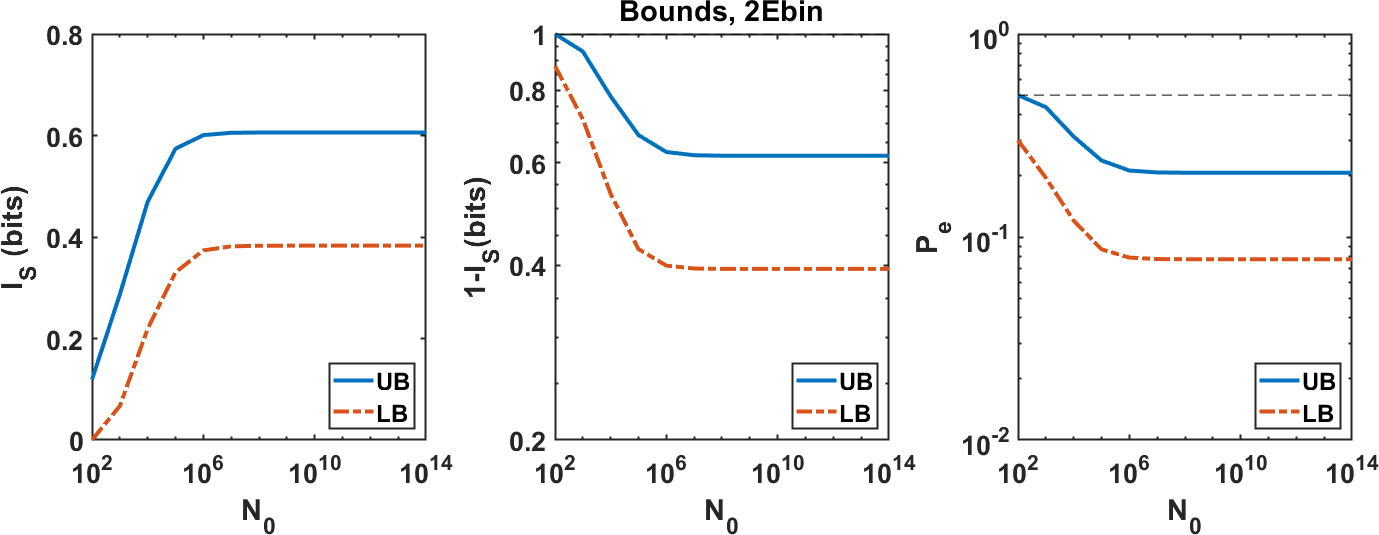}
	\includegraphics[width=\linewidth]{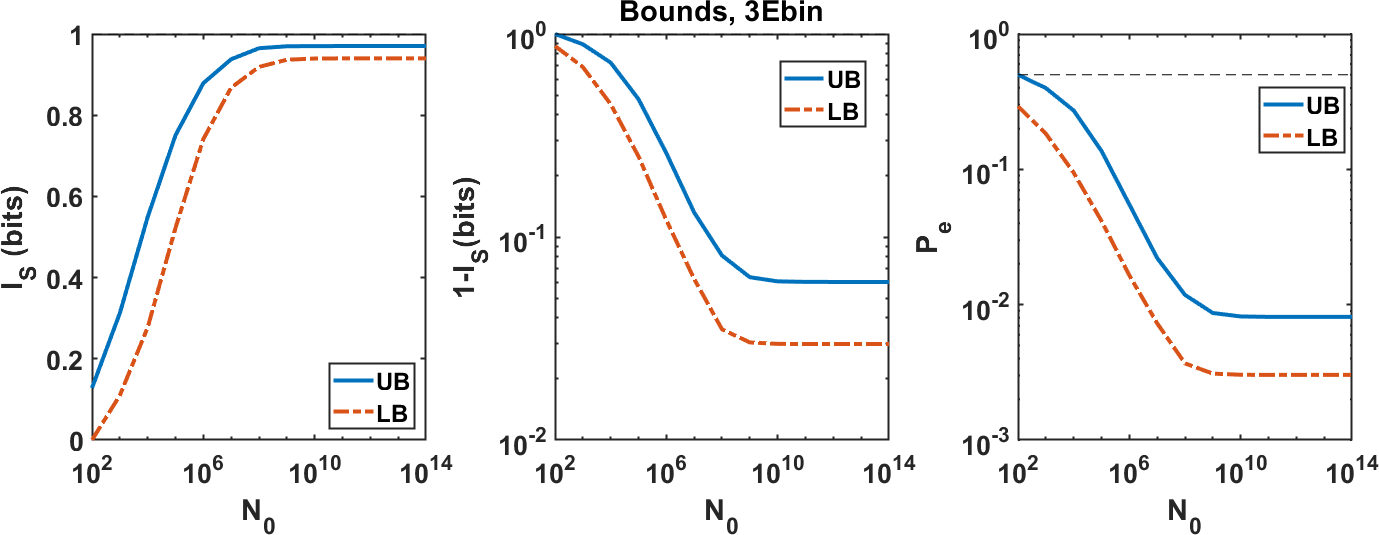}
	\caption{Upper bound (blue) and lower bound (red) on $\smi$, $\ent(C)-\smi$ and detection error probability $\pe$ for one (top), two (middle) and three (bottom) energy bins. The maximum possible value of $\pe$, which is 0.5 for equal prior, is plotted in black dashed line.}
	\label{fig:SMI_separate1}
\end{figure}

%\begin{figure}[H]
%\centering
%	\includegraphics[width=\linewidth]{fig/Ebin2_SMI.png}
%	\caption{Upper bound (blue) and lower bound (red) on $\smi$, $\ent(C)-\smi$ and $\pe$ for 2 energy bin. The maximum possible value of each metric is plotted in black dashed line.}
%	\label{fig:SMI_separate2}
%\end{figure}
%
%\begin{figure}[H]
%\centering
%	\includegraphics[width=\linewidth]{fig/Ebin3_SMI.png}
%	\caption{Upper bound (blue) and lower bound (red) on $\smi$, $\ent(C)-\smi$ and $\pe$ for 3 energy bin. The maximum possible value of each metric is plotted in black dashed line.}
%	\label{fig:SMI_separate3}
%\end{figure}

\indent Figure~\ref{fig:SMI_separate1} presents bounds on $\smi$ and $\pe$ as functions of source photon budget $N_0$ for one, two and three energy bins, respectively. The upper bound (blue solid line) and lower bound (red dashed line) of $\smi$, $\ent(C)-\smi$, and $\pe$ are shown from left to right in each figure. Note that for binary classification tasks with equal prevalence, $\ent(C)=1$. The $\ent(C)-\smi$ is plotted, because it is in the same order of magnitude with $\pe$ when $\pe\ll1$. 

\indent For a binary classification task with equal prevalence of the two classes, $\smi\in [0,1]$ and $\pe\in [0, 0.5]$. The task-specific performance improves when $\smi$ increases and $\pe$ decreases. When the source photon budget $N_0$ increases, both upper bound and lower bound of $\smi$ increases, which means that, as source count increases, the imaging system captures more information for the binary classification task. However, both $\smi$ and $\pe$ saturates at high photon regime, which is due to material variability. 

\begin{figure}[H]
\centering
	\includegraphics[width=0.5\linewidth]{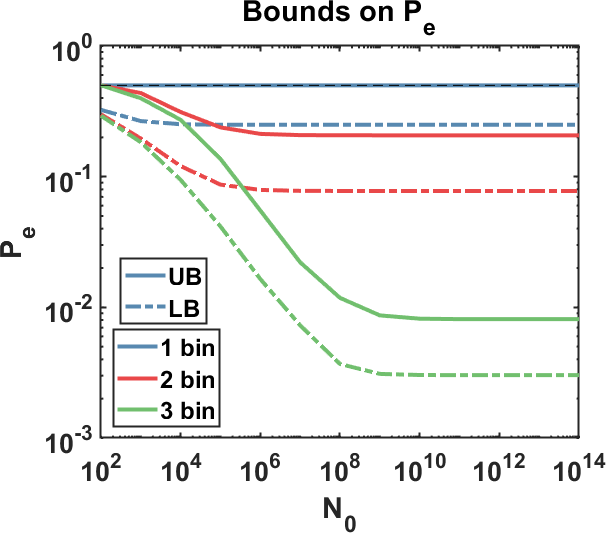}
	\caption{Bounds on detection error rate $P_e$ for 1, 2, and 3 energy bins.}
	\label{fig:PE_combined}
\end{figure}

\indent To show $\pe$ for all three energy binning occasions, Figure~\ref{fig:PE_combined} presents the bounds on $\pe$ for 1, 2, and 3 bins. At a fixed source count $N_0$, bounds on $\pe$ decrease as the number of energy bins increases. In fact, for systems with only one energy bin, the upper bound on $\pe$ is 0.5 and the lower bound saturates at around 0.25, which means that binary classification with such a system is close to random guess. The specific number of bounds on $\pe$ changes with the material library and other simulation setups, but the general trend that more energy bins provides higher task-specific information is valid.

\subsection{Three measurement models}

\begin{figure}
\centering
	\includegraphics[width=0.49\linewidth]{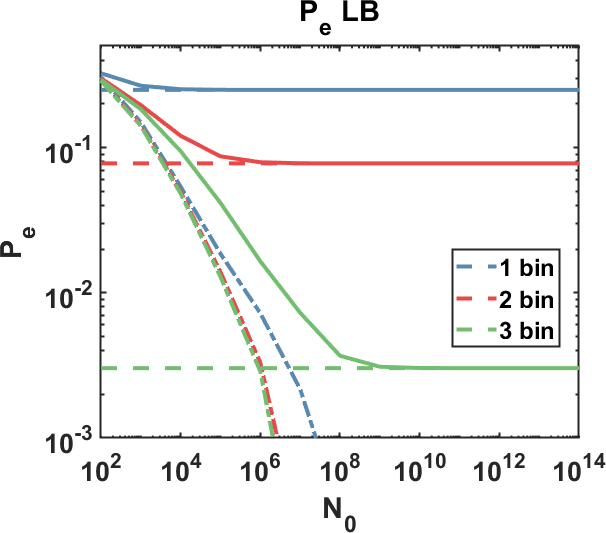}
	\includegraphics[width=0.49\linewidth]{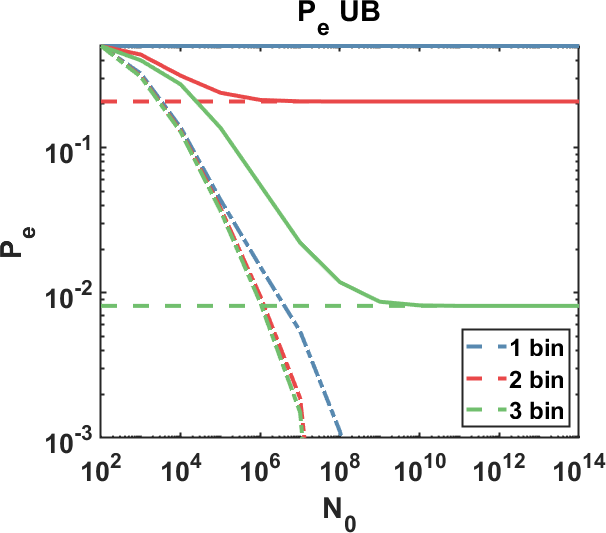}
	\caption{Lower bound (left) and upper bound (right) on $\pe$ as functions of source photon count for measurement model that considers only the shot noise (dotted lines), that considers only the energy-correlated material variation (dashed line) and that considers both the shot noise and the energy-correlated material variation. Results for detectors that have one, two and three energy bins are indicated in blue, red and green.}
\label{fig:pe_result}
\end{figure}

\indent Figure~\ref{fig:pe_result} shows the lower bound and the upper bound of $\pe$ for three measurement models as a function of source photon number ($N_0$). The three measurement modes consist of a model that considers only the shot noise (approximated as Gaussian, dot-dashed lines), a model that considers only the energy-correlated material variation (dashed lines), and a model that considers both the shot noise and the energy-correlated material variation (solid lines). Results for detectors with 1, 2 and 3 energy bins are presented in different colors. At low count region, the system performance is limited by the quantum noise as the bounds on $\pe$ of the combined model is similar to that of the model considering only quantum noise; at high count region, the system performance is limited by the material variation as the bounds on $\pe$ of the combined model approaches that of the model considering only material variation. 

\subsection{Energy-correlated model vs.\ energy-uncorrelated model}

\begin{figure}
\centering
	\includegraphics[width=0.8\linewidth]{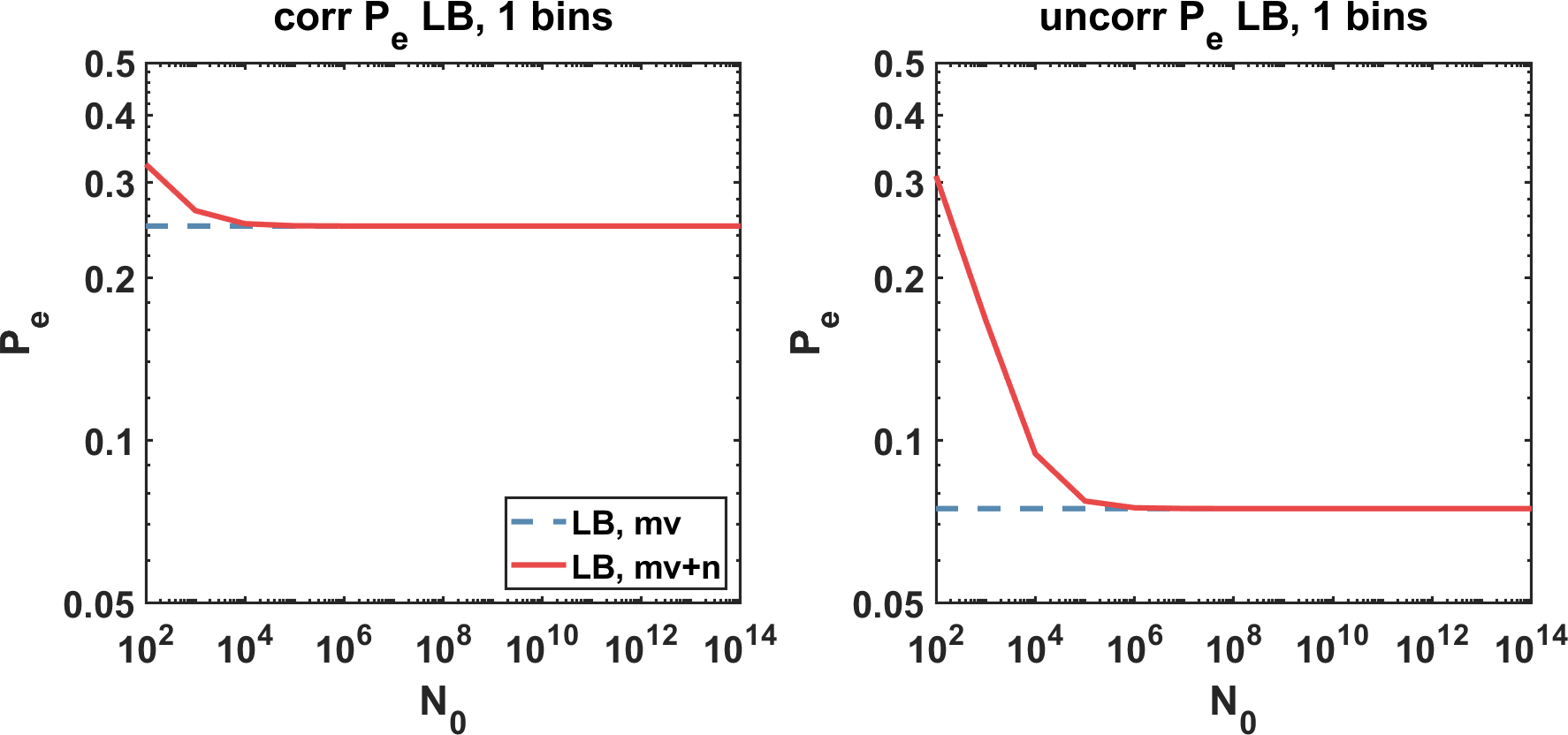}
	\includegraphics[width=0.8\linewidth]{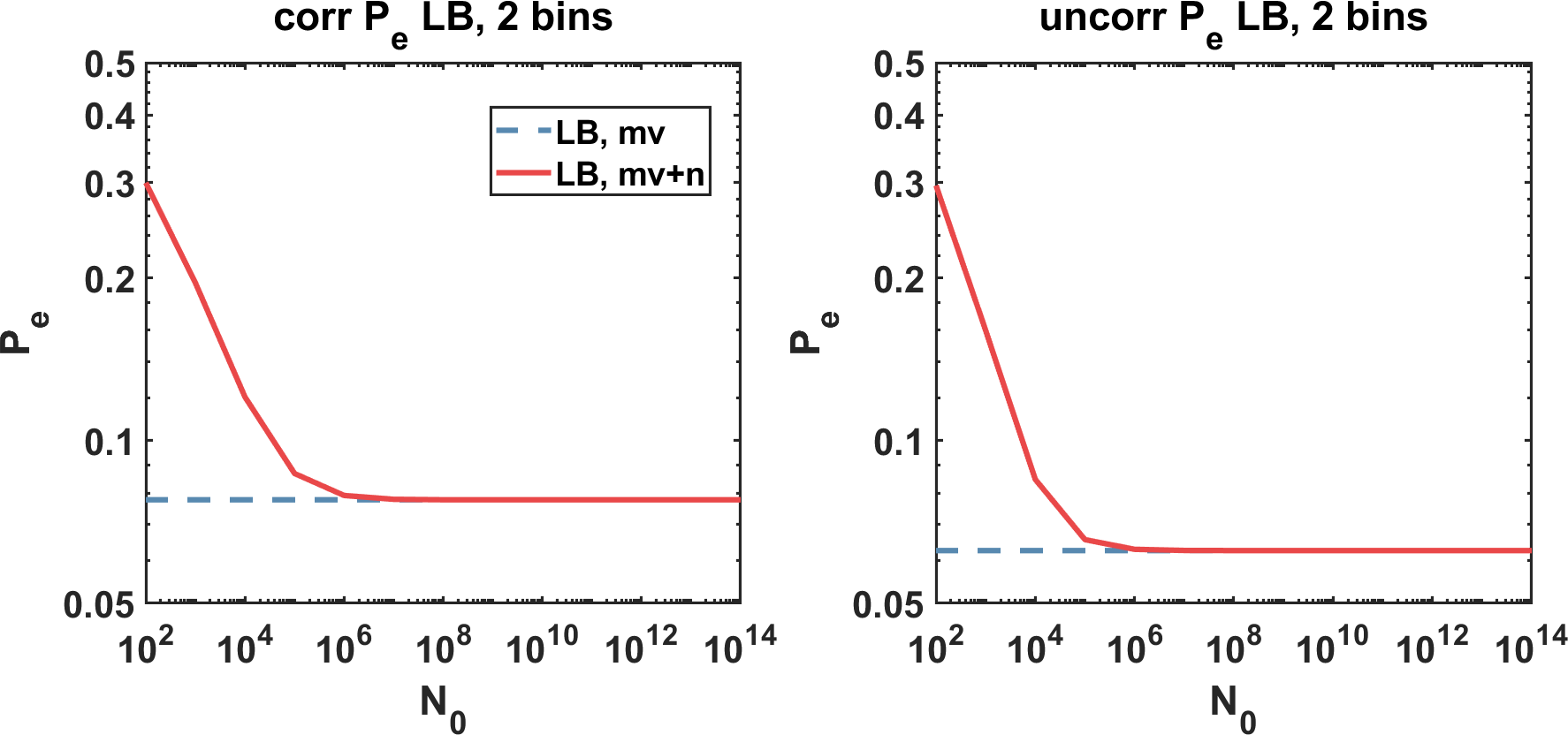}
	\includegraphics[width=0.8\linewidth]{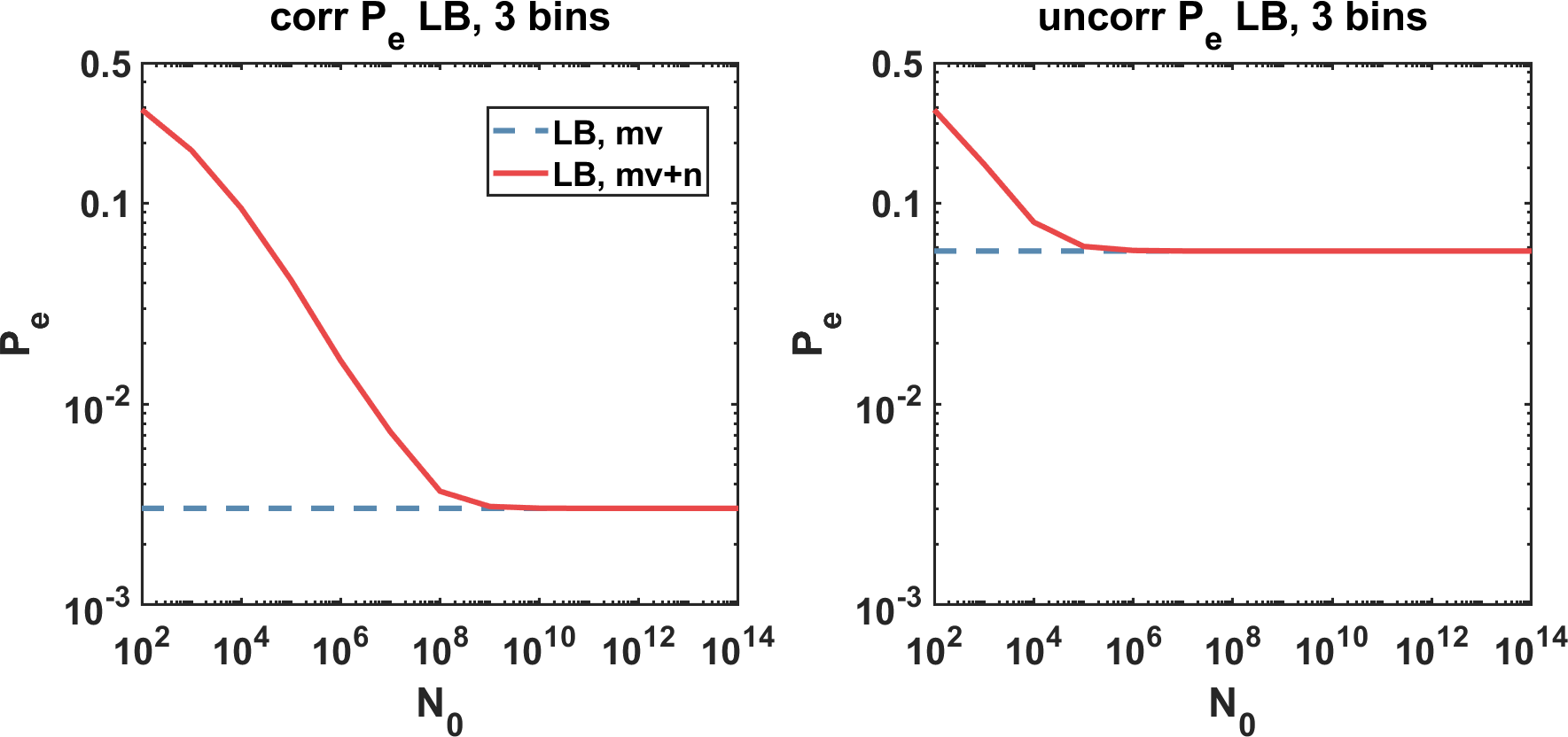}
	\caption{Comparison of the energy-correlated model with energy-uncorrelated model in terms of lower bound on $\pe$. The material-variation-only model (blue lines) and the full measurement model (red dashed lines) are considered.}
	\label{fig:pe_uncorr}
\end{figure}

\indent We compare the energy-correlated models with the energy-uncorrelated models in Figure~\ref{fig:pe_uncorr}. From top to bottom, results of detectors with 1, 2 and 3 energy bins are presented. In each figure, the results calculated from the material-variation model are also presented for reference (red dashed line). When there is one energy bin, the bound on $\pe$ of the energy-uncorrelated model is lower than that of the correlated model. When the number of energy bins equals to 2, the bound on $\pe$ of the uncorrelated model is close to that of the correlated model. When the number of energy bins equals to 3, the bound on $\pe$ of the uncorrelated model is much higher than that of the correlated model. 

\indent The difference in bound on $\pe$ calculated from the energy-uncorrelated model and the energy-correlated model is due to the difference in the material variation considered in the two models, as indicated by the difference in the red-dashed lines. When energy correlation in the data is considered, $\Sigma_J$, which is the covariance matrix of $J(E)$, is calculated based on Equation~(\ref{eq:cov_IE}); in contrary, when energy correlation is not considered, the off-diagonal elements of $\Sigma_J$ are ignored. Since the energy-binning process is an integration over energy, each element of $\Sigma_{J_d}$ is the sum of a block matrix in $\Sigma_J$. For example, when there is one energy bin, $\Sigma_{J_d}$ is the sum of all elements in $\Sigma_J$ when energy correlation is considered; while in the uncorrelated model, $\Sigma_{J_d}$ is the sum of only the diagonal elements of $\Sigma_J$. Therefore, $\Sigma_{J_d}$ of the energy-correlated model is larger than that of the uncorrelated model for the one energy bin scenario. 

\indent When there are more than one energy bin, the comparison is more complex and we introduce a method to quantitatively compare the  material variability model with energy correlation and that without energy correlation. The size of the material variability can be quantified by the volume of the noise bubble induced by the material variability. If we define the volume of the noise bubble as the volume of a $M$-dimensional ellipsoid determined by the covariance matrix $\Sigma_{J_d}$, where $M$ is the number of energy bins. The eigenvector of the covariance matrix from singular value decomposition provides the principle axes of the ellipsoid and the square root of the eigenvalues are half the length of the principle axes. For example, when energy correlation is not considered, $\Sigma_{J_d}$ is diagonal with each element equals to the variance of detected photon count in the corresponding energy bin; the square root of an eigenvalue is the standard deviation of the detected photon count; and the volume of the ellipsoid determined by the covariance matrix is the product of the standard deviation in each energy bin. 

\indent To compare the material variability estimated by the energy-correlated model and uncorrelated model, we can calculate the ratio of the two volumes determined by the two ellipsoids. Denote this ratio as $r = V_{corr}/V_{uncorr}$. In the dataset used in Figure~\ref{fig:pe_uncorr}, $\log_{10}(r)$ is $1.00\pm0.07$ (mean $\pm$ standard deviation) for one energy bin, $0.98\pm0.45$ for two energy bins and $-1.08\pm1.05$ for three energy bins. For the set of bags we studied, the uncorrelated model always underestimates the material variability for one energy bin; and with three energy bins, the uncorrelated model often severely overestimates the material variability.

%%%%%%%%%%%%%%%%%%%%% Discussion %%%%%%%%%%%%%%%%%%%%%%%
\section{Discussion} \label{sec:discussion}

\indent To incorporate material variation, a linear perturbation of the beer's law is used in the derivation from Equation~\ref{eq:beers_law} to Equation~\ref{eq:beers_law_linear}. This perturbation is equivalent to approximating the sum of correlated log-normal random variables with a normal distribution, which has been justified in Section~\ref{sec:meas_model} for small material variations. One direction of future work is to examine other approximations of the sum of log-normal random variables and developing a measurement model that accommodates large material variations. Such a model could further reduce the number of materials in the material library and hence reduce the number of objects needed in a study.

\indent At low count region, the measurement model should be applied with caution, due to the break down of using Gaussian distribution to approximate the statistics of the shot noise. In comparison, Masoudi et.\ at.\ \cite{masoudi2018x} incorporated the Poisson distribution in an energy-uncorrelated model. Their model performances better than our model, when shot noise is dominating and material variation is not important. However, many X-ray systems are not photon starving and the effect of material variation should be considered. To address the concern about the accuracy of the Poisson-Gaussian approximation, we provide an error analysis in Appendix~B. 

\indent We assumed the attenuation profiles, $\mu(E)$, follow normal distributions and derived that the measurement data $\bm g$ following normal distributions under the following two assumptions: (1) The material variation $\Delta\tau$ is small, and (2) the distribution of the shot noise can be approximated by a Gaussian. Although our derivation relied heavily on the normality of data, the mean and covariance matrix of the data $\bm g$ are independent to the shape of the distributions, as long as the above assumption (1) is still valid. 

\indent The main goals of this work include: presenting the framework, justifying the approximations used in the measurement model, and studying the effect of energy correlation. In future work, we will incorporate spatial correlation and other sources of variations, such as source fluctuation and detector variation. 

\section{Conclusion} \label{sec:conclusion}
\indent In this work, we have presented an energy-correlated X-ray measurement model that incorporates both material variation and shot noise. Energy correlations are inherent in a material’s attenuation profile and affect the system performance. Therefore, it is important to consider the energy correlations. We successfully modeled the shot noise and energy-correlated material variation by a multivariate Gaussian model. Furthermore, under the assumption of no spatial correlation, we provided analytical forms for TSI metrics, including bounds on $\smi$ and bounds on $\pe$, for binary classification tasks. Spatial correlation of the X-ray measurement data is not explored in this work and will be a focus of future study. 
   
\section*{Funding}
The authors gratefully acknowledge the support of the US Department of Homeland Security. The research for this project was conducted under contract with the U.S Department of Homeland Security (DHS) Science and Technology Directorate (S\&T), contract HSHQDC-16-C-B0014. The opinions contained herein are those of the contractors and do not necessarily reflect those of DHS S\&T.

%\section*{Acknowledgments}

\section*{Disclosures}
The authors declare that there are no conflicts of interest related to this article.

%%%%%%%%%%%%%%%%%%%%%%% References %%%%%%%%%%%%%%%%%%%%

\end{document}

\title{Appendix A: Bounds on Shannon mutual information}

%\author{Yijun Ding,\authormark{1,*} and Amit Ashok,\authormark{1,2}}
%
%\address{\authormark{1}College of Optical Sciences, Tucson, AZ, USA\\
%\authormark{2}Department of Electrical and Computer Engineering, Tucson, AZ, USA
%}
%
%\email{\authormark{*}dingy@email.arizona.edu} %% email address is required

%%%%%%%%%%%%%%%%%% abstract %%%%%%%%%%%%%%%%
% [use \begin{abstract*}...\end{abstract*} if exempt from copyright]

\begin{abstract}
We provide two new bounds on Shannon mutual information ($\smi$) for the task of binary classification with mixture data. More specifically, the upper bound is based solely on pair-wise Kullback-Leibler (KL) divergence, and the lower bound is based solely on pair-wise Bhattacharyaa distance (BD).
\end{abstract}

%%%%%%%%%%%%%%%%%%%%  Introduction  %%%%%%%%%%%%%%%%%%%%%%%
\section{Upper bound on $\smi$ based on pair-wise KL divergence}
\indent When the number of objects in the two classes are equal ($K_1 = K_2$), each object in category 1 can form a pair with an object in category 2. A second upper bound on $\smi$ can be derived (see appendix).
\begin{equation}
\begin{split}
\smi(\bm g; C) &\leq \ent(C) - \sum_{i=1}^{K} a_i\ent(\bm g_i) +\sum_{i=1}^{K_1} (a_{i,1} + a_{i,2}) [\ent(\bm g_{i,pair})-\ent_{i}(C)] \\
&\leq \ent(C) - \sum_{i=1}^{K} a_i\ent(\bm g_i) +\sum_{i=1}^{K_1} (a_{i,1} + a_{i,2}) [\hat\ent_{KL}(\bm g_{i,pair}) - \ent_{i}(C)] ,
\label{eq:bounds_Is2}
\end{split}
\end{equation}
where $a_i,1$ and $a_i,2$ are the probability of the two objects in the $i_th$ object pair, $\bm g_{i,pair}$ is the measurement data when the two objects in the $i_th$ pair are imaged and $H_i(C)$ is the Shannon entropy of the class label for the $i_{th}$ object pair. If we define $r_{i,t} = {a_{i,t}}/({a_{i,t} + a_{i,nt}})$ and $r_{i,nt} = {a_{i,nt}}/({a_{i,t} + a_{i,nt}})$, we can calculate the entropy by $\ent_{i}(c) = -r_{i,t}\ln(r_{i,t}) - r_{i,nt}\ln(r_{i,nt})$.

\section{Lower bound on $\smi$ based on pair-wise Bhattacharyaa distance}
Denote $P_c$ as the probability of the class label $C=c$. If we consider the measured data as a mixture of data conditioned on class label $C$, we can lower bound $\smi$ by
\begin{equation}
\begin{split}
	\smi(\bm g; C) &\geq - \sum_{c=1}^2 P_c\ln\left[ P_c + P_{c'} e^{-\bd{\prob(\bm g|C=c)}{\prob(\bm g|C=c')}}\right] \\
	&= \ent(C) - \sum_{c=1}^2 P_c\ln\left[ 1 + \frac{P_{c'}}{P_c} e^{-\bd{\prob(\bm g|C=c)}{\prob(\bm g|C=c')}}\right],
	\end{split}
\end{equation}
where $c'\in\{1,2\}$ and $c'\neq c$. The inequality has been proved in Kolchinsky et.\ al.\ \cite{kolchinsky2017estimating} while studying Shannon entropy for mixture distributions. When the conditional data can be further considered as mixture distributions, the Bhattacharyaa distance (BD) between two conditioned distribution $\prob(\bm g|C=1)$ and $\prob(\bm g|C=2)$ can be bounded by pair-wise BD.
\begin{equation}
\begin{split}
e^{-\bd{\prob(\bm g|C=1)}{\prob(\bm g|C=2)}} &= {\int \mathrm{dg}\, \prob(\bm g|C=1)\,\prob(\bm g|C=2)} \\
& = {\int\mathrm{dg}\,\left[\sum_{i,C=1}\frac{w_i}{P_1}\prob_i \right]^{0.5}\left[\sum_{j,C=2}\frac{w_j}{P_2}\prob_j\right]^{0.5}} \\
&\leq {\int\mathrm{dg}\,\sum_{i,C=1}\left[\frac{w_i}{P_1}\prob_i\right]^{0.5} \sum_{j,C=2}\left[\frac{w_j}{P_2}\prob_j\right]^{0.5}} \\
&= \sum_{i,C=1}\sum_{j,C=2}\left(\frac{w_iw_j}{P_1P_2}\right)^{0.5}\,e^{-\bd{\prob_i}{\prob_j}},
\end{split}
\end{equation}
where the inequality stems from $\sqrt{a+b}\leq \sqrt{a} + \sqrt{b}$ for $a\geq0$ and $b\geq 0$. Therefore, the lower bound on $\smi$ based on pair-wise BD is:
\begin{equation}
\smi(\bm g; C) \geq \ent(C) -\sum_{c=1}^2 P_c\ln\left[ 1+\frac{P_{c'}^{0.5}}{P_c^{1.5}}\,\sum_{i\in\{c\}}\sum_{j\in\{c'\}} (w_iw_j)^{0.5}e^{-\bd{\prob_i}{\prob_j}} \right]
\end{equation}

%\section*{References}
%\bibliographystyle{plain}
%% begin .bib file
\bibliography{math}

\title{Appendix B: Approximating a Poisson distribution with a normal distribution}

%\author{Yijun Ding,\authormark{1,*} and Amit Ashok,\authormark{1,2}}
%
%\address{\authormark{1}College of Optical Sciences, Tucson, AZ, USA\\
%\authormark{2}Department of Electrical and Computer Engineering, Tucson, AZ, USA
%}
%
%\email{\authormark{*}dingy@email.arizona.edu} %% email address is required

%%%%%%%%%%%%%%%%%% abstract %%%%%%%%%%%%%%%%
\begin{abstract*}
We provide an error bound analysis for approximating a Poisson distribution with a normal distribution. 
\end{abstract*}

\section{Distribution Approximation}

To simplify notations for the proofs, we consider a random variable $k$ that follows a Poisson distribution with mean $\lambda$
\begin{equation}
\Prob(k|\lambda) = \frac{\lambda^k e^{-\lambda}}{k!},
\end{equation}
whose mean and variance both equal to $\lambda$.

\indent The denominator $k!$ can be write as Stirling's formula multiplied by a $k$-related constant:
\begin{equation}
	k! = \sqrt{2\pi k}k^ke^{-k}\cdot e^{r_k},
	\label{eq:stir}
\end{equation}
for $k=1,2,\cdots$, where $r_k$ is bounded by \cite{robbins1955remark}
\begin{equation}
	\frac{1}{12k+1}<r_k<\frac{1}{12k}.
\end{equation}
The value of $r_k$ is small for large $k$. For example, when $k\geq 9$, $r_k$ is less than 0.01. Therefore, $e^{r_k}$ is often approximated by 1 when $k$ is large.

\indent Now let $x=k=\lambda(1+\delta)$ where $\lambda\gg1$. Since the mean and variance $\overline{k} = \text{var}(k)= \lambda$, $\delta$ is on the order of $\lambda^{-0.5}$. Therefore, when $\lambda\gg1$, $\delta\ll1$. We are concerned with large values of $k$, in which case the discrete $\Prob(k)$ goes over to a continuous probability density function for variable $x$. With Equation~\ref{eq:stir}, we find
\begin{equation}
	\prob(x|\lambda) = \frac{e^{\lambda\delta}(1+\delta)^{-\lambda(1+\delta)-0.5}}{\sqrt{2\pi\lambda}}\cdot e^{r_x}
 \label{eq:poiss_gauss1}
\end{equation}

Since
\begin{equation}
\begin{split}
	\ln{(1+\delta)^{-\lambda(1+\delta)}} &=  -\lambda(1+\delta)[\delta-\delta^2/2+\delta^3/3+O(\delta^4)] \\
								   &= -\lambda\delta-\lambda\delta^2/2  + \lambda\delta^3/6 + O(\delta^2)), \\
\end{split}
\end{equation}
and
\begin{equation}
r_x = O(\frac{1}{12\lambda(1+\delta)}) = O(\delta^2),
\end{equation}
where $O$ is the order of the function. 

We have 
\begin{equation}
\begin{split}
	\prob(x|\lambda) &= \frac{e^{-\lambda\delta^2/2}}{\sqrt{2\pi\lambda}}\cdot \frac{e^{\frac{\lambda\delta^3}{6}+O(\delta^2)}}{\sqrt{1+\delta}} \\
		  &=\mathcal N(x; \lambda, \lambda)\cdot[1+\frac{-3\lambda+(x-\lambda)^2}{6\lambda}\delta+O(\delta^2)]
\end{split}
\end{equation}
where $\mathcal N(x; \lambda, \lambda) =  \exp[{-(x-\lambda)^2/(2\lambda)}]/\sqrt{2\pi\lambda}$ is a normal distribution with mean and variance both equal to $\lambda$. When $\delta\ll1$ and $x\gg1$, the Poisson distribution with mean $\lambda$ can be approximated by a normal distribution $\mathcal N(x; \lambda, \lambda)$ and the error of this approximation is
\begin{equation}
	\frac{\prob(x|\lambda) - \mathcal N(x; \lambda, \lambda)}{\prob(x|\lambda)} = \frac{-3\lambda+(x-\lambda)^2}{6\lambda}\delta +O(\delta^2).
	\label{eq:poiss_error1}
\end{equation}
When $\lambda\geq100$, the absolute value of error percentage defined by Equation~\ref{eq:poiss_error1} is less than $3.4\%$ for $x=\lambda\pm\sqrt{\lambda}$.

Another expression of Poisson distribution is
\begin{equation}
\begin{split}
	\prob(x|\lambda) &= \frac{e^{-\lambda\delta^2/[2(1+\delta)]}}{\sqrt{2\pi x}}\cdot e^{-\frac{\lambda\delta^3}{3}+O(\delta^2)}\\
		  &=\frac{e^{-(x-\lambda)^2/(2x)}}{\sqrt{2\pi x}}\cdot[1-\frac{(x-\lambda)^2}{3\lambda}\delta+O(\delta^2)]
\end{split}
\end{equation}
When $\delta\ll1$ and $x\gg1$, the Poisson distribution with mean $\lambda$ can be approximated with $f(x) =\exp[-(x-\lambda)^2/(2x)]/\sqrt{2\pi x}$.

\section{Error Analysis} 
The error of approximating a Poisson distribution with a normal distribution is
\begin{equation}
	\frac{\prob(x|\lambda) - f(x)}{\prob(x|\lambda)} = -\frac{(x-\lambda)^2}{3\lambda}\delta +O(\delta^2).
\end{equation}
When $\lambda\geq100$, the absolute value of error percentage is less than $3.6\%$ for $x=\lambda-\sqrt{\lambda}$ and less than $3.1\%$ for $x=\lambda+\sqrt{\lambda}$.

%
%An equivalent of Equation~(\ref{eq:poiss_gauss1}) is
%\begin{equation}
%	\prob(x|\lambda)  \approx \frac{e^{\lambda\delta}(1+\delta)^{-\lambda(1+\delta)}}{\sqrt{2\pi x}}.
%\end{equation}
%Note that, up to this step, we only used Stirling's formula, which has an error of $0.83\%$ for $x=10$ and the accuracy increases with increasing $x$. Further assume $\delta\ll1$:
%
%As a result,
%\begin{equation}
%	\prob(x|\lambda)  \approx \frac{e^{-(x-\lambda)^2/(2x)}}{\sqrt{2\pi x}}.
%\end{equation}

\bibliography{math}

\title{Appendix C: Combining shot noise and material variation}

%\author{Yijun Ding,\authormark{1,*} and Amit Ashok,\authormark{1,2}}
%
%\address{\authormark{1}College of Optical Sciences, Tucson, AZ, USA\\
%\authormark{2}Department of Electrical and Computer Engineering, Tucson, AZ, USA
%}
%
%\email{\authormark{*}dingy@email.arizona.edu} %% email address is required

%%%%%%%%%%%%%%%%%% abstract %%%%%%%%%%%%%%%%
\begin{abstract*}
We provide a derivation for combining shot noise and the material variation in our measurement model. 
\end{abstract*}

\section{Derivation}

Continue from the paper, we have the distribution of the mean photon flux
\begin{equation}
\prob(\bm J_{d} | \bm J_{d,0}) \approx \mathcal N(\bm J_d; \bm J_{d,0},\Sigma_{J_d}),
\end{equation}
where $\Sigma_{J_d}$ is given in the main paper and it is related with the material variation. The approximation stems from the linear perturbation of the Beer's law. 

For a given mean photon flux, the distribution of the shot noise can be described by
\begin{equation}
\prob(\bm g | \bm J_{d}) \approx \mathcal N(\bm g; \bm J_{d},{\text diag}(\bm J_{d}))
\end{equation}
where $\text{diag}(\bm J_d)$ returns a square diagonal matrix with the elements of vector $\bm J_d$ on the diagonal. The approximation sign is from approximating a Poisson distribution with a Gaussian distribution.

Combing shot noise and the material variation, the distribution of the measurement data follows
\begin{equation}
	\begin{split}
		\prob(\mathbf g| \mathbf J_{d0}, \Sigma_{J_d}) &= \int_0^\infty \mathrm{d\mathbf J_d}~ \prob(\mathbf g|\mathbf J_d) \,\prob(\mathbf J_d| \mathbf J_{d0}, \Sigma_{J_d})\\
		&\approx \int_0^\infty \mathrm{d\mathbf J_d}~ \mathcal N(\mathbf g; \mathbf J_d, , \text{diag}(\mathbf J_{d}))\, \mathcal N(\mathbf J_d; \mathbf J_{d0}, \Sigma_{J_d})\\
		&\approx \int_0^\infty \mathrm{d\mathbf J_d}~ \mathcal N(\mathbf J_d; \mathbf g, , \text{diag}(\mathbf g)) \,\mathcal N(\mathbf J_d; \mathbf J_{d0}, \Sigma_{J_d})\\
		&\approx \int_{-\infty}^\infty \mathrm{d\mathbf J_d}~ \mathcal N(\mathbf J_d; \mathbf g , \text{diag}(\mathbf g)) \,\mathcal N(\mathbf J_d; \mathbf J_{d0}, \Sigma_{J_d})\\
		&=\mathcal N(\mathbf g; \mathbf J_{d0},\, \Sigma_{J_d} + \text{diag}(\mathbf g )) \\
		&\approx \mathcal N(\mathbf g; \mathbf J_{d0},\, \Sigma_{J_d} + \text{diag}(\mathbf J_{d0})).
	\end{split}
\end{equation}

The second approximation $ N(\mathbf g; \mathbf J_d, , \text{diag}(\mathbf I_{d}))\approx  N(\mathbf J_d; \mathbf g , \text{diag}(\mathbf g))$ can be approved through the two approximations of a Poisson distribution discussed in Appendix B. The third step of the approximations is true when each element of $J_{d0}\gg0$ with respect to $\Sigma_{J_d}$. Lastly, the fourth step of the approximation requires that the variance is small.

\bibliography{math}